\documentclass{article}
\usepackage{amsmath}
\usepackage{cite}
\usepackage{graphicx}
\usepackage{dcolumn}

\begin{document}

\date{}
\title{Sheared potentials and travelling nodes}
\author{Francisco M. Fern\'{a}ndez\thanks{%
fernande@quimica.unlp.edu.ar} \\
INIFTA, DQT, Sucursal 4, C. C. 16, \\
1900 La Plata, Argentina}
\maketitle

\begin{abstract}
When a sheared potential is deformed in such a way that the distance between
the classical turning points remains constant the eigenvalues of the
Schr\"{o}dinger equation oscillate with respect to the potential parameter
responsible for the deformation. We show that such an oscillation is
intimately related to the passing of the nodes of the corresponding
eigenfunctions through the origin. We illustrate this effect by means of the
split harmonic oscillator and the split linear potential.
\end{abstract}

\section{Introduction}

\label{sec:intro}

The so called shared potentials have received some interest for
several years. For example, Gosh and Hasse\cite{GH81} resorted to
the split harmonic oscillator to show that not all classical
harmonic oscillators are quantum harmonic oscillators. Osypowski
and Olsson\cite{OO87} studied asymmetric potentials for which the
classical period is independent of the energy and chose the split
harmonic oscillator as one of the examples. Stillinger and
Stillinger\cite{SS89} showed that the uncorrected semiclassical
approximation applied to pseudoharmonic oscillators misses several
significant qualitative features of the exact spectrum and the
split oscillator was one of their examples. Dorignac\cite{D05}
calculated the first semiclassical corrections to the WKB approach
and resorted to the split harmonic oscillator as a suitable
example. Asorey et al\cite{ACMP07} characterized isoperiodic
potentials showing that they are connected by some transformations
and chose the split harmonic oscillator as one of the examples.
Ant\'{o}n and Brun\cite{AB08} discussed several classical models
with periods that are independent of the energy, one of which is
the split harmonic oscillator.

In a recent paper, Oliveira-Cony et al\cite{ORFS25} solved the
Schr\"{o}dinger equation for two sheared potentials: the split harmonic
oscillator and the split linear potential. They tried to explain some
features of the spectrum by focusing on the eigenfunctions. Part of their
analysis was based on the classical force and the work made by an external
agent to produce the deformation of the potential.

In this paper we analyse the two models chosen by Oliveira-Cony et al\cite
{ORFS25} but pay attention to the migration of the nodes of the wavefunction
as the potential is deformed. In section~\ref{sec:HO} we consider the split
harmonic oscillator, section~\ref{sec:linear} is devoted to the split linear
potential; finally, in section~\ref{sec:conclusions} we summarize the main
results and draw conclusions.

\section{Split harmonic oscillator}

\label{sec:HO}

Oliveira-Cony et al\cite{ORFS25} discussed a quantum-mechanical model for a
particle of mass $m$ moving in one dimension under the split harmonic
potential
\begin{equation}
V(x)=\left\{
\begin{array}{c}
\frac{\kappa \nu ^{2}}{(2\nu -1)^{2}}x^{2},\;x<0, \\
\kappa \nu ^{2}x^{2},\;x\geq 0.
\end{array}
\right.  \label{eq:V(x)_HO}
\end{equation}
The main feature of this potential is that the distance between the left and
right classical turning points $x_{-}(\nu )<0<x_{+}(\nu )$ is independent of
$\nu $: $x_{+}(\nu )-x_{-}(\nu )=\mathrm{const}$.

The eigenvalues and eigenfunctions of the Schr\"{o}dinger equation are $%
E_{n}(\nu )$ and $\psi _{n\,\nu }(x)$, respectively, where the quantum
number $n=0,1,\ldots $ is the number of nodes of $\psi _{n\,\nu }(x)$ in $%
-\infty <x<\infty $. These nodes are located between $x_{-}$ and $x_{+}$ for
any value of $\nu $.

When $\nu =1$ we have the usual harmonic oscillator with eigenvalues
\begin{equation}
E_{n}(1)=\hbar \omega \left( n+\frac{1}{2}\right) ,\;\omega =\sqrt{\frac{%
2\kappa }{m}}.  \label{eq:E_n(1)}
\end{equation}
When $\nu =1/2$ the problem reduces to solve the Schr\"{o}dinger equation in
$0<x<\infty $ with the boundary conditions $\psi (0)=0$, $\psi \left(
x\rightarrow \infty \right) =0$ and the eigenvalues are
\begin{equation}
E_{n}(1/2)=\frac{\hbar \omega }{2}\left( 2n+\frac{3}{2}\right) ,
\label{eqE_n(1/2)}
\end{equation}
so that $E_{n}(1)<E_{n}(1/2)$. However, $E_{n}(\nu )$ does not increase
monotonously when $\nu $ decreases from $1$ towards $1/2$ when $n>1$ but
exhibits interesting oscillations. The classical argument of Oliveira-Cony
et al does not account for those oscillations. In what follows we focus on
the behaviour of the nodes of the eigenfunctions that provide much more
information.

As $\nu $ decreases the classical turning points move to the right and,
consequently, the nodes of $\psi _{n\,\nu }(x)$ also move to the right and
eventually one of the nodes in $x<0$ crosses the origin towards $x>0$.
Suppose that for a given value of $\nu =\nu _{ij}$ there are $i$ nodes in $%
x<0$, $j$ nodes in $x>0$ and one node exactly at $x=0$. Obviously, $n=i+j+1$%
. If we solve the Schr\"{o}dinger equation for $x<0$ with the boundary
condition $\psi (0)=0$ we obtain
\begin{equation}
E_{n_{i}}(\nu )=\hbar \omega \frac{\nu }{2\nu -1}\left( n_{i}+\frac{1}{2}%
\right) ,\;n_{i}=2i+1.  \label{eq:E_i}
\end{equation}
Analogously, for $x>0$ we have
\begin{equation}
E_{n_{j}}(\nu )=\hbar \omega \nu \left( n_{j}+\frac{1}{2}\right)
,\;n_{j}=2j+1.  \label{eq:E_j}
\end{equation}
From $E_{n_{i}}(\nu )=E_{n_{j}}(\nu )$ we obtain $\nu =\nu _{ij}$, where
\begin{eqnarray}
\nu _{ij} &=&\frac{2(i+j)+3}{4j+3}=\frac{2n+1}{4(n-i)-1},  \nonumber \\
E_{n}\left( \nu _{ij}\right)  &=&\hbar \omega \left( i+j+\frac{3}{2}\right)
=\hbar \omega \left( n+\frac{1}{2}\right) =E_{n}(1).  \label{eq:nu,E(nu)}
\end{eqnarray}
Note that $1-\nu _{ij}=\frac{2(j-i)}{4j+3}\geq 0$ (because $j\geq i$) and
that $\nu _{ij}-\frac{1}{2}>0$. Besides, $\nu _{ij}=1$ only when $i=j$ and $n
$ is odd.

When $\nu =1/2$ the $n$ nodes are located in $x>0$ and there is an
additional node at $x=0$ due to the impenetrable wall at this point;
consequently,
\begin{equation}
E_{n}\left( \frac{1}{2}\right) =\frac{\hbar \omega }{2}\left( 2n+\frac{3}{2}%
\right) >E_{n}(1).  \label{eq:E_n(1/2)}
\end{equation}
We conclude that $E_{n}(\nu )$ increases from $E_{n}(1)$ to $E_{n}(1/2)$ in
an oscillatory way and reaches the value $E_{n}(1)$ every time a node of $%
\psi _{n\,\nu }(x)$ is located at $x=0$ during its migration from $x<0$
towards $x>0$. It is worth noting that when one of the nodes is located at $%
x=0$ the energy $E_{n}\left( \nu _{ij}\right) $ does not depend on $i$ and $%
j $ separately but on their sum $i+j$. Besides, from present
analytical expressions we obtain $E_{n}(1/2)/E_{n}(1)$ that
accounts for the numerical results in figure~5 of Oliveira-Cony et
al.

\section{Split linear potential}

\label{sec:linear}

The second example is is the split linear potential
\begin{equation}
V(x)=\left\{
\begin{array}{c}
-\frac{\kappa \nu }{2\nu -1}x,\;x<0, \\
\kappa \nu x,\;x\geq 0.
\end{array}
\right.   \label{eq:V(x)_lin}
\end{equation}
In this case the Schr\"{o}dinger equation can be solved in terms of the Airy
function $Ai(z)$ and the eigenvalues, assuming that there is a zero at $x=0$%
, can be expressed in terms of the zeros $0>a_{1}>a_{2}\ldots $ of $Ai(z)$%
\cite{AS72}. A straightforward calculation shows that
\begin{equation}
E_{i}=-\frac{a_{i}}{2^{1/3}}\left( \frac{\hbar ^{2}\kappa ^{2}}{m}\right)
^{1/3}\left( \frac{\nu }{2\nu -1}\right) ^{2/3},\;E_{j}=-\frac{a_{j}}{2^{1/3}%
}\left( \frac{\hbar ^{2}\kappa ^{2}}{m}\right) ^{1/3}\nu ^{2/3}.
\label{eq:Ei,Ej}
\end{equation}
From $E_{i}(\nu )=E_{j}(\nu )$ we obtain
\begin{equation}
\nu _{ij}=\frac{1}{2}\left[ 1+\left( \frac{a_{i}}{a_{j}}\right)
^{3/2}\right] ,  \label{eq:nu_(ij)}
\end{equation}
and
\begin{equation}
E_{ij}=\left( \frac{\hbar ^{2}\kappa ^{2}}{m}\right) ^{1/3}\frac{\left(
|a_{i}|^{3/2}+|a_{j}|^{3/2}\right) ^{2/3}}{2}.  \label{eq:E_(ij)}
\end{equation}
Note that $\nu _{ij}>\frac{1}{2}$ and $\nu _{ij}\leq 1$ because $i\leq j$.
In this case the number of zeros is $n=i+j-1$; therefore, $\nu _{ij}=1$ when
$i=j$ and $n$ is odd.

Table~\ref{Tab:En_linear} shows several eigenvalues $E_{n}\left( \nu
_{ij}\right) =E_{ij}$ in units of $\left( \frac{\hbar ^{2}\kappa ^{2}}{m}%
\right) ^{1/3}$. We appreciate that for a given value of $n$ the eigenvalues
$E_{n}\left( \nu _{ij}\right) $ change slightly with the values of $i$ and $j
$ that satisfy $n=i+j-1$. This slight variation with the individual values
of $i$ and $j$ becomes less prominent as $i$ and $j$ increases. The reason
is that the zeros of $Ai(z)$ behave asymptotically as
\begin{equation}
a_{i}\sim -\left[ \frac{2\pi }{2}\left( i-\frac{1}{4}\right) \right]
^{2/3},\;i\gg 1,  \label{eq:ai_asympt}
\end{equation}
so that
\begin{equation}
E_{n}\left( \nu _{ij}\right) \sim \left( \frac{\hbar ^{2}\kappa ^{2}}{m}%
\right) ^{1/3}\frac{\left( 6\pi \right) ^{2/3}}{8}\left( 2i+2j-1\right)
^{2/3},\;i,j\gg 1.  \label{eq:E_n_asymp}
\end{equation}
In the semiclassical limit the spectrum of the split linear potential
behaves as the spectrum of the split harmonic oscillator in that $%
E_{n}\left( \nu _{ij}\right) $ depends on the sum $i+j$. As in the preceding
example we conclude that $E_{n}(1/2)>E_{n}(1)$ and $E_{n}(\nu )$ oscillates
as $\nu $ decreases from $\nu =1$ to $\nu =1/2$ in such a way that $%
E_{n}(\nu )$ reaches a value close to $E_{n}(1)$ every time a zero
is located at $x=0$; that is to say, for $\nu =\nu _{ij}$. The
main difference
is that $E_{n}\left( \nu _{ij}\right) $ is close, but not identical, to $%
E_{n}(1)$.

\section{Conclusions}

\label{sec:conclusions}

The deformation of a potential-energy function under the requirement that
the distance between classical turning points remains constant produces an
oscillation of the energy eigenvalues. This oscillation is due to the
passage of a zero of the wavefunction through $x=0$. This conclusion is just
a conjecture derived from the analysis of two simple, exactly solvable
examples.

\begin{table}[tbp]
\caption{Eigenvalues $E_n(\nu_{ij})$ for the split linear potential}
\label{Tab:En_linear}
\begin{center}
\small
\begin{tabular}{rrrll}
\multicolumn{1}{c}{$i+j$} & \multicolumn{1}{c}{$i$} & \multicolumn{1}{c}{$j$}
& \multicolumn{1}{c}{$\nu_{ij}$} & \multicolumn{1}{c}{$E_{ij}$} \\
2 & 1 & 1 & 1 & 1.855757081 \\
3 & 1 & 2 & 0.7162760442 & 2.597461596 \\
4 & 1 & 3 & 0.6378135787 & 3.246651172 \\
4 & 2 & 2 & 1 & 3.244607624 \\
5 & 1 & 4 & 0.6011062347 & 3.836630657 \\
5 & 2 & 3 & 0.8186057411 & 3.834331402 \\
6 & 1 & 5 & 0.5798356533 & 4.384362798 \\
6 & 2 & 4 & 0.7337434899 & 4.382063620 \\
6 & 3 & 3 & 1 & 4.381671239 \\
7 & 1 & 6 & 0.5659576940 & 4.899820070 \\
7 & 2 & 5 & 0.6845688774 & 4.897577389 \\
7 & 3 & 4 & 0.8668224701 & 4.897065861 \\
8 & 1 & 7 & 0.5561894538 & 5.389474508 \\
8 & 2 & 6 & 0.6524849742 & 5.387300034 \\
8 & 3 & 5 & 0.7896508969 & 5.386747623 \\
8 & 4 & 4 & 1 & 5.386613780 \\
9 & 1 & 8 & 0.5489410219 & 5.857822816 \\
9 & 2 & 7 & 0.6299021674 & 5.855715801 \\
9 & 3 & 6 & 0.7393004182 & 5.855151291 \\
9 & 4 & 5 & 0.8948107334 & 5.854960865 \\
10 & 1 & 9 & 0.5433488546 & 6.308148112 \\
10 & 2 & 8 & 0.6131448055 & 6.306104199 \\
10 & 3 & 7 & 0.7038603682 & 6.305539692 \\
10 & 4 & 6 & 0.8261801521 & 6.305322798 \\
10 & 5 & 5 & 1 & 6.305263006 \\
11 & 1 & 10 & 0.5389035130 & 6.742939434 \\
11 & 2 & 9 & 0.6002164960 & 6.740953468 \\
11 & 3 & 8 & 0.6775624084 & 6.740394399 \\
11 & 4 & 7 & 0.7778733377 & 6.740164761 \\
11 & 5 & 6 & 0.9130842002 & 6.740074630
\end{tabular}
\end{center}
\end{table}

\end{document}